\documentclass[aip,jcp,reprint]{revtex4-1}

\usepackage{graphicx}
\usepackage{amssymb,amsfonts,amsmath}
\usepackage{epsf}
\usepackage{subfigure}
\usepackage{epstopdf}
\usepackage{mathrsfs}
\usepackage{color}

\newcommand{\be}{\begin{equation}}
\newcommand{\ee}{\end{equation}}
\newcommand{\bea}{\begin{eqnarray}}
\newcommand{\eea}{\end{eqnarray}}

\begin{document}

\title{Fluctuating chemohydrodynamics and \\ the stochastic motion of self-diffusiophoretic particles}

\author{Pierre Gaspard}
\email{gaspard@ulb.ac.be}
\affiliation{Center for Nonlinear Phenomena and Complex Systems, Universit{\'e} Libre de Bruxelles, Code Postal 231, Campus Plaine, B-1050 Brussels, Belgium}
\author{Raymond Kapral}
\email{rkapral@chem.utoronto.ca }
\affiliation{Chemical Physics Theory Group, Department of Chemistry, University of Toronto, Toronto, Ontario M5S 3H6, Canada}

\begin{abstract}

The propulsion of active particles by self-diffusiophoresis is driven by asymmetric catalytic reactions on the particle surface that generate a mechanochemical coupling between the fluid velocity and the concentration fields of fuel and product in the surrounding solution. Because of thermal and molecular fluctuations in the solution, the motion of micrometric or submicrometric active particles is stochastic. Coupled Langevin equations describing the translation, rotation, and reaction of such active particles are deduced from fluctuating chemohydrodynamics and fluctuating boundary conditions at the interface between the fluid and the particle. These equations are consistent with microreversibility and the Onsager-Casimir reciprocal relations between affinities and currents, and provide a thermodynamically consistent basis for the investigation of the dynamics of active particles propelled by diffusiophoretic mechanisms.
\end{abstract}

\maketitle

\section{Introduction}

Nature often makes use of molecular machines that convert chemical energy supplied by their environments into directed motion that is then exploited to carry out various transport and other biological functions.\cite{alberts-cell} One need not rely on Nature to construct small machines, and synthetic self-propelled nano- and micro-motors use catalytic chemical reactions on a portion of the motor to achieve directed motion. Such motors have been made and are the focus of much interest because of the myriad of potential applications that make use of their small size and ability to carry out active transport.\cite{PKOSACMLC04,FBAMO05,W13,WDAMS13,SSK15,DWDYMS15,YDBS15} These active particles are driven by gradients of concentrations, electrochemical potentials, or temperature generated by the  surface reactions, and operate in nonequilibrium systems where energy transduction from reaction to motion is induced by mechanochemical coupling through diffusiophoresis, electrophoresis, or thermophoresis.\cite{GLA07,K13,CRRK14}

Because of their micrometric or submicrometric sizes, active particles are subjected to thermal and molecular fluctuations in the solution in which they reside, so that a stochastic description of the system is required to study their dynamics. In this regard, a challenging issue is how to bridge the gap between the fluctuating chemohydrodynamics describing the fluid and the stochastic movements of the active particle while remaining consistent with the principles of nonequilibrium thermodynamics.\cite{P67,N79,C85,KB08,H69,GM84}

The purpose of the present paper is to address this issue and set up a framework to deduce overdamped Langevin equations ruling the stochastic translation, rotation, and reaction of an active particle self-propelled by diffusiophoresis, starting from a fluctuating continuous-medium description.  The key element in this analysis is the need to understand the interplay between the boundary conditions for the fluid velocity and solute concentration fields at the interface with the active particle.  The boundary conditions express the coupling between the interfacial irreversible processes generating the mechanochemical coupling.  These irreversible processes are the surface reaction, the diffusiophoretic effect, and the interfacial friction due to slip velocity.  To be consistent with microreversibility, the linear response coefficients describing these interfacial processes must satisfy the Onsager-Casimir reciprocal relations.\cite{O31a,O31b,C45,W67,BAM76}  Since the mechanochemical coupling is generated by diffusiophoresis, there exists a reciprocal effect back onto the concentration fields and the reaction, which has consequences for the coupled stochastic equations ruling the motion and reaction of the particle. Because of this reciprocal effect, the reaction rate depends on the mechanical force exerted on the particle.  The inclusion of this reciprocal effect is essential in order to obtain the mechanochemical fluctuation theorem that governs the stochastic motion of the chemically-propelled motor.\cite{GK17}

The plan of this paper is the following.  In Sec.~\ref{FluctChemoHydro}, the fluctuating chemohydrodynamics formulation is presented for a solution containing a Janus particle with a catalytic surface where an interfacial reaction takes place.  The frequency-dependent force, torque, and reaction rate of the particle are deduced from these boundary conditions in Sec.~\ref{sec:GLangevin}.  In Sec.~\ref{Low-freq}, the low-frequency limit is considered in order to obtain analytical expressions for these quantities for a spherical Janus particle composed of catalytic and noncatalytic hemispheres. In this way, the diffusiophoretic force is expressed in terms of diffusiophoretic constants and the reaction rate.  In Sec.~\ref{Langevin}, the coupled overdamped Langevin equations, as well as the associated Fokker-Planck equation, are deduced for the stochastic motion and reaction of the Janus motor and their implications are studied.  The conclusions are given in  Sec.~\ref{conclusion}, which summarizes the results and presents a perspective on the work.

\section{Fluctuating chemohydrodynamics with surface reactions}
\label{FluctChemoHydro}

Fluctuating thermodynamics methods are well known and provide a way to incorporate thermal fluctuations in continuum descriptions of the dynamics.\cite{LL80Part1,LL80Part2,OS06} In the linear regime, close to thermodynamic equilibrium, the fluctuation-dissipation theorem provides a systematic method to set up the stochastic differential equations describing the random motion of the variables that are used to describe the dynamics of the system. To implement this scheme, the different irreversible processes $\{\alpha\}$ are identified and the thermodynamic entropy production rate is written as a linear combination of products of their affinities $\{A_\alpha\}$ and noiseless currents $\{\langle J_\alpha\rangle\}$:~\cite{P67,N79,H69,GM84,C85,KB08}
\be
\frac{1}{k_{\rm B}}\frac{d_{\rm i}S}{dt} = \sum_\alpha A_\alpha \langle J_\alpha\rangle \geq 0 \, ,
\label{entrprod1}
\ee
where $k_{\rm B}$ is Boltzmann's constant and $\langle\cdot\rangle$ denotes the statistical average over the fluctuations.  Phenomenological linear relations are established between the currents and the affinities
\be
\langle J_\alpha\rangle = \sum_\beta L_{\alpha\beta}  A_\beta
\ee
in terms of linear response coefficients $L_{\alpha\beta}$, so that the entropy production rate takes the quadratic form:
\be
\frac{1}{k_{\rm B}}\frac{d_{\rm i}S}{dt} = \sum_{\alpha,\beta} L_{\alpha\beta} A_\alpha A_\beta \geq 0 \, .
\label{entrprod2}
\ee
As a consequence of microreversibility, the linear response coefficients obey the Onsager-Casimir reciprocal relations $L_{\alpha\beta}=\epsilon_\alpha\epsilon_\beta L_{\beta\alpha}$ where $\epsilon_\alpha=\pm1$ when $A_\alpha$ is even or odd under time reversal.\cite{O31a,O31b,C45,W67,BAM76,GM84,H69}  Accordingly, only the coefficients $L_{\alpha\beta}$ that couple processes with the same parity under time reversal contribute to the entropy production rate~(\ref{entrprod2}).

Fluctuating currents are obtained by adding a noise term $\delta J_\alpha(t)$ to the mean currents:
\be
J_\alpha = \sum_\beta L_{\alpha\beta}  A_\beta + \delta J_\alpha(t) \, .
\label{J-alpha}
\ee
The fluctuating quantities $\delta J_\alpha(t)$ are assumed to be Gaussian white noise processes characterized by
\bea
&&\langle\delta J_\alpha(t)\rangle = 0  \qquad \mbox{and}\nonumber\\
&&\langle\delta J_\alpha(t)\, \delta J_\beta(t')\rangle = (L_{\alpha\beta}+L_{\beta\alpha})\, \delta(t-t') \, ,
\label{Gaussian}
\eea
which vanish if $L_{\alpha\beta}$ couples processes with opposite parities under time reversal.

Making the assumption of local thermodynamic equilibrium, the same principles may be used to construct the stochastic equations for spatially extended systems.\cite{LL80Part2,BM74,OS06}

The general stochastic thermodynamics method outlined above will now be applied to describe Janus motors propelled by a diffusiophoretic mechanism. A Janus motor is spherical particle with catalytic and noncatalytic hemispheres, and we suppose that it is suspended in a multi-component fluid containing solute species, labeled by the index $k$, that interact with the motor through short range intermolecular potentials $u_k$. We further assume that the reversible reactions ${\rm A} \rightleftharpoons{\rm B}$ occur on the catalytic hemisphere, and call species ${\rm A}$  the fuel and ${\rm B}$  the product. These chemical reactions produce inhomogeneous $c_{\rm A}$ and $c_{\rm B}$ concentration fields in the motor vicinity that lead to a body force on the motor. If no external forces act on the system and momentum is conserved, fluid flows arise in the surrounding medium that are responsible for motor propulsion. The forms that the stochastic equations for the fluid velocity and concentration fields take in the solution and on the surface are given below.

\subsection{Stochastic equations in the bulk phases}

The hydrodynamic and diffusive processes in the solution surrounding the catalytic Janus particle are described by the coupled Navier-Stokes and diffusion equations. The fluctuating Navier-Stokes equation ruling the velocity field $\bf v$ is given by
\be
\rho\left(\partial_t{\bf v} + {\bf v}\cdot\pmb{\nabla}{\bf v}\right) = -{\rm div}\, {\boldsymbol{\mathsf P}} \, ,
\ee
where $\rho$ is the mass density and ${\boldsymbol{\mathsf P}}$ is the pressure tensor.\cite{LL80Part2,BM74,OS06} The fluid is assumed to be incompressible,
\be
\pmb{\nabla}\cdot{\bf v} = 0,
\label{div.v=0}
\ee
so that the mass density remains uniform. In this case, the pressure tensor is related to the gradients of the velocity field by the phenomenological linear relations:
\be
P_{ij} = P \, \delta_{ij} -\eta \left( \partial_i v_j+\partial_j v_i\right) + \pi_{ij} \, ,
\label{Pij}
\ee
where $P$ is the hydrostatic pressure, $\eta$ the shear viscosity, and $\pi_{ij}$ are Gaussian white noise fields characterized by
\bea \label{pi-correl}
&&\langle \pi_{ij}({\bf r},t)\rangle = 0 \qquad\mbox{and}\\
&&\langle \pi_{ij}({\bf r},t)\, \pi_{kl}({\bf r}',t')\rangle = 2 k_{\rm B}T \eta \left(\delta_{ik}\delta_{jl}+\delta_{il}\delta_{jk}\right)\nonumber\\
&&\qquad\qquad\qquad\qquad\qquad\times \delta({\bf r}-{\bf r}') \, \delta(t-t'),\nonumber
\eea
$T$ denoting the temperature, in accord with the general relations given in Eq.~(\ref{Gaussian}).\cite{LL80Part2,BM74,OS06}

The fluctuating diffusion equations for the concentration fields $c_k$ of the different solute species $k=1,2,...,s$,
\be
\partial_t \, c_k + \pmb{\nabla}\cdot{\bf j}_k = 0,
\label{diff-eq-1}
\ee
contain the fluctuating current densities,
\be
{\bf j}_k = c_k {\bf v} - D_k \pmb{\nabla}c_k + \pmb{\eta}_k \, ,
\label{diff-eq-2}
\ee
where $D_k$ is the molecular diffusivity of species $k$ and $\pmb{\eta}_k$ are Gaussian white noise fields satisfying~\cite{OS06}
\bea
&&\langle \pmb{\eta}_k({\bf r},t)\rangle = 0 \quad \mbox{and}  \\
&&\langle \pmb{\eta}_k({\bf r},t)\,   \pmb{\eta}_{k'}({\bf r}',t')\rangle =2 D_k c_k \delta_{kk'} \delta({\bf r}-{\bf r}') \, \delta(t-t')\, {\boldsymbol{\mathsf 1}} \, , \nonumber
\eea
with ${\boldsymbol{\mathsf 1}}$ denoting the $3\times 3$ identity matrix.  Moreover, the noise terms on the pressure and current densities are uncorrelated: $\langle \pi_{ij} ({\bf r},t)\,  \pmb{\eta}_{k'}({\bf r}',t')\rangle=0$.

Inside a solid Janus particle of radius $R$, the velocity and concentration fields take the following values:\cite{MB74}
\be
{\bf v}({\bf r},t) = {\bf V}(t) + \pmb{\Omega}(t)\times \left[ {\bf r}-{\bf R}(t)\right] \quad\mbox{and} \quad c_k({\bf r},t)=0,
\ee
for $\Vert{\bf r}-{\bf R}(t)\Vert <R$ and $k=1,2,...,s$, where ${\bf R}(t)$ is the position of the center of mass of the Janus particle, ${\bf V}(t)$ its velocity, and $\pmb{\Omega}(t)$ its angular velocity.

\subsection{Stochastic equations at the interface}

Several irreversible processes take place at the interface between the fluid and the Janus particle.  First, there is a frictional force along the interface associated with the partial slip of the velocity field between the fluid and the solid particle.  Second, the reaction ${\rm A}\rightleftharpoons{\rm B}$ takes place on the catalytic hemisphere of the Janus particle with local rate,
\be
w= \kappa_+ c_{\rm A}-\kappa_- c_{\rm B} \, ,
\label{ws}
\ee
where $\kappa_\pm$ are rate coefficients. Third, a diffusiophoretic force is exerted by the diffusing species on the Janus particle.  All these processes can be described by boundary conditions on the velocity and concentration fields by extending interface nonequilibrium thermodynamics\cite{W67,BAM76,K77,B86} to stochastic processes.\cite{BAM77}

We expect there to be a coupling between the tangential components of the pressure tensor $\boldsymbol{\mathsf P}$ and the surface current density of species $k$ through their respective affinities:
\bea
{\bf n}\cdot{\boldsymbol{\mathsf P}}\cdot{\boldsymbol{\mathsf 1}}_{\bot} &=& -\frac{L_{\rm vv} }{T}\, {\bf v}_{\rm slip} - \sum_k \frac{L_{{\rm v}k}}{T}\, \pmb{\nabla}_{\bot}\mu_k^{\rm s} + \mbox{noise} ,  \qquad \label{BC3} \\
{\bf j}_k^{\rm s} &=& -\frac{L_{k{\rm v}}}{T}\, {\bf v}_{\rm slip} - \sum_l \frac{\tilde L_{kl}}{T}\, \pmb{\nabla}_{\bot}\mu_l^{\rm s} + \mbox{noise}, \label{BC2}
\eea
where $\bf n$ is a unit vector normal to the interface and oriented towards the fluid, ${\boldsymbol{\mathsf 1}}_{\bot}\equiv{\boldsymbol{\mathsf 1}}-{\bf n}{\bf n}$, ${\bf v}_{\rm slip}$ is the slip velocity between the fluid and the solid, $\pmb{\nabla}_\bot$ denotes the tangential gradient, and $\mu_k^{\rm s}$ is the surface chemical potential of species $k$.  The quantity $\lambda=L_{\rm vv}/T$ is the coefficient of sliding friction.\cite{BB13}  The coefficients $\tilde L_{kl}$ in Eq.~(\ref{BC2}) are related to the surface diffusion coefficients of the adsorbates, if they exist.
Diffusiophoresis is characterized by the coefficients $L_{{\rm v}k}$.  In order to satisfy microreversibility, the linear response coefficients must obey the Onsager-Casimir reciprocal relations,
\be
\tilde L_{kl} = \tilde L_{lk}  \quad \mbox{and}\quad \ L_{k{\rm v}}=-L_{{\rm v}k} \, ,
\label{ORR-vect}
\ee
because ${\bf v}_{\rm slip}$ is odd under time reversal, while $\pmb{\nabla}_{\bot}\mu_l^{\rm s}$ is even.
The latter relations show that diffusiophoresis has a reciprocal effect back onto the surface current density~(\ref{BC2}).

Equations~(\ref{BC3}) and~(\ref{BC2}) represent boundary conditions for the  velocity and concentration fields, respectively. In particular, we deduce from Eq.~(\ref{BC3}) the following boundary condition on the tangential component of the velocity field:
\bea
{\bf v}_{\rm slip} &\equiv& {\boldsymbol{\mathsf 1}}_{\bot}\cdot\left\{ {\bf v}({\bf r},t) -{\bf V}(t) - \pmb{\Omega}(t)\times\left[{\bf r}-{\bf R}(t)\right]\right\} \nonumber\\
&=&{\boldsymbol{\mathsf 1}}_{\bot}\cdot\biggl\{ b\left[\pmb{\nabla}{\bf v}({\bf r},t) +\pmb{\nabla}{\bf v}({\bf r},t)^{\rm T}\right]\cdot{\bf n} -{\bf v}^{\rm s}_{\rm fl}({\bf r},t) \nonumber\\
&& \qquad \qquad\qquad - \sum_k  b_k \pmb{\nabla}c_k({\bf r},t)\biggr\} \label{v-bc2}
\eea
for $\Vert{\bf r}-{\bf R}(t)\Vert =R$.  Here $b=\eta/\lambda$ is the slip length, $b_k$ the diffusiophoretic constants
\be
b_k \equiv k_{\rm B}  \, \frac{L_{{\rm v}k}}{\lambda c_k} = \frac{k_{\rm B}T}{\eta} \left( K_k^{(1)} + b\, K_k^{(0)}\right) \, ,
\label{b_k}
\ee
given in terms of the quantities
\be
K_k^{(n)} \equiv \int_R^{R+\delta} dr \, (r-R)^n \, \left[ {\rm e}^{-\beta u_k(r)}-1 \right] \, , \label{K-dfn}
\ee
where $\delta$ is the finite range of the intermolecular potentials $u_k(r)$ in the direction $r$ that is radial from the center of mass of the Janus particle, and $\beta=(k_{\rm B}T)^{-1}$.\cite{A89,AP91,AB06} The fluctuating interfacial velocity field, ${\bf v}^{\rm s}_{\rm fl}({\bf r},t)$, is  a Gaussian white noise process satisfying
\bea
&&\langle{\bf v}^{\rm s}_{\rm fl}({\bf r},t)\rangle = 0 \qquad\mbox{and} \nonumber\\
&& \delta^{\rm s}({\bf r},t)\, \langle{\bf v}^{\rm s}_{\rm fl}({\bf r},t)\, {\bf v}^{\rm s}_{\rm fl}({\bf r}',t')\rangle\, \delta^{\rm s}({\bf r}',t')\nonumber\\
&&\qquad = 2 k_{\rm B}T \, \frac{b}{\eta}\, {\boldsymbol{\mathsf 1}}_{\bot} \delta^{\rm s}({\bf r},t) \, \delta({\bf r}-{\bf r}') \, \delta(t-t') ,
\eea
where $\delta^{\rm s}({\bf r},t)$ is the interfacial Dirac distribution.\cite{BAM76}  This fluctuating velocity field is related by ${\bf v}^{\rm s}_{\rm fl}({\bf r},t)={\bf F}_{\rm R}({\bf r},t)/\lambda$ to the fluctuating force per unit area considered in Ref.~\onlinecite{BAM77}.  The expression~(\ref{v-bc2}) for the slip velocity that includes the diffusiophoretic term with the constants~(\ref{b_k}) in the presence of partial slip was obtained previously.\cite{AB06}  Here we assume that the reaction has negligible effect on the diffusiophoretic constants.  In addition to the boundary condition~(\ref{v-bc2}), the normal component of the velocity field obeys~\cite{ABM75}
\be
 {\bf n}\cdot{\bf v}({\bf r},t) = {\bf n}\cdot{\bf V}(t) \label{v-bc1}
\ee
for $\Vert{\bf r}-{\bf R}(t)\Vert =R$.

The boundary conditions for the concentration fields are determined by the surface reaction rate~(\ref{ws}), as well as the reciprocal effect of diffusiophoresis.  This effect is a consequence of microreversibility and the Onsager-Casimir reciprocal relations in the coupling between interfacial solute transport and the slip velocity, in analogy with the cross effects related to thermal slip.\cite{W67,BAM76} Indeed, the boundary conditions on the concentration fields can be expressed as
\be
{\bf n}\cdot(c_k\, {\bf v}-D_k \, \pmb{\nabla}c_k)_R =\nu_{k} \, w  - \Sigma_k^{\rm s} + \mbox{noise}\, ,
\label{BC1bis}
\ee
in terms of is the stoichiometric coefficient $\nu_k$ which is positive for the product, $\nu_{\rm B}=1$, and negative for the reactant, $\nu_{\rm A}=-1$, and a possible sink into a boundary layer with excess surface density $\Gamma_k$ for species $k$
\be
\Sigma_k^{\rm s} = \partial_t\Gamma_k +\pmb{\nabla}_{\bot}\cdot(\Gamma_k{\bf v}^{\rm s}+{\bf j}_k^{\rm s}) ,
\ee
where ${\bf v}^{\rm s}$ is the surface velocity and ${\bf j}_k^{\rm s}$ is the surface current density~(\ref{BC2}).  With the reaction rate~(\ref{ws}), these boundary conditions are thus of the form:
\bea
&&{\bf n}\cdot\left(c_{\rm A}{\bf v}-D_{\rm A}\pmb{\nabla}c_{\rm A}\right)_R = -\left(\kappa_+ c_{\rm A}-\kappa_- c_{\rm B}\right)_{R} \nonumber\\
&&\qquad\qquad \qquad\qquad\qquad - \xi^{\rm s}({\bf r},t)-\Sigma_{\rm A}^{\rm s}  , \label{bc-cA}\\
&&{\bf n}\cdot\left(c_{\rm B}{\bf v}-D_{\rm B}\pmb{\nabla}c_{\rm B}\right)_R = +\left(\kappa_+ c_{\rm A}-\kappa_- c_{\rm B}\right)_{R} \nonumber\\
&&\qquad\qquad \qquad\qquad\qquad +\xi^{\rm s}({\bf r},t) -\Sigma_{\rm B}^{\rm s}   ,  \label{bc-cB}
\eea
where $\xi^{\rm s}({\bf r},t)$ is the interfacial noise associated with the surface reaction~(\ref{ws}) and satisfies,
\bea
&&\langle\xi^{\rm s}({\bf r},t)\rangle = 0 \qquad\mbox{and} \nonumber\\ && \delta^{\rm s}({\bf r},t)\, \langle\xi^{\rm s}({\bf r},t)\, \xi^{\rm s}({\bf r}',t')\rangle\, \delta^{\rm s}({\bf r}',t')\nonumber\\ && \qquad= \left(\kappa_+ c_{\rm A}+\kappa_- c_{\rm B}\right)  \delta^{\rm s}({\bf r},t) \, \delta({\bf r}-{\bf r}') \, \delta(t-t') ,
\eea
analogous to that for bulk phase reactions.\cite{OS06}

These boundary conditions will be used in the following section to determine the effects of the surrounding solution on the Janus motor.

\section{Frequency-dependent force, torque and reaction rate} \label{sec:GLangevin}

The force exerted on the Janus particle by the fluid is determined by the surface integral of the pressure tensor at the interface ${\cal S}(t)$ between the fluid and the Janus particle and, if present, an external force ${\bf F}_{\rm ext}$.  As a consequence, Newton's equation for the Janus particle is given by
\be
m\frac{d{\bf V}}{dt} = -\int_{{\cal S}(t)} {\boldsymbol{\mathsf P}}({\bf r},t)\cdot{\bf n} \, dS + {\bf F}_{\rm ext},
\label{Langevin-Eq-0}
\ee
where $m=\int_{{\cal V}(t)}\rho_{\rm sol} \, d{\bf r}$ is the mass of the Janus particle and $\rho_{\rm sol}$ its mass density.\cite{MB74,BM74}

In a similar manner, a torque is exerted by the fluid on the Janus particle so that the angular velocity obeys the equation,
\be
{\boldsymbol{\mathsf I}}\cdot\frac{d\pmb{\Omega}}{dt} = -\int_{{\cal S}(t)} {\bf r}\times \left[{\boldsymbol{\mathsf P}}({\bf r},t)\cdot{\bf n}\right]  dS + {\bf T}_{\rm ext},
\ee
where the inertia tensor ${\boldsymbol{\mathsf I}}$ of the Janus particle has the components $I_{ij}=\int_{{\cal V}(t)}\rho_{\rm sol} \big(\Delta{\bf r}^2\, \delta_{ij}-\Delta r_i\Delta r_j\big) d{\bf r}$ with $\Delta{\bf r}\equiv{\bf r}-{\bf R}(t)$, and ${\bf T}_{\rm ext}$ is an external torque.\cite{H75,F76a,F76b,BAM77}

The overall reaction rate of the Janus particle is given by
\be
W= \int_{{\cal S}(t)} dS \left( \kappa_+ \, c_{\rm A} - \kappa_- \, c_{\rm B}  \right)_{R} \, ,
\label{W}
\ee
where the surface integral is carried out over the catalytic hemisphere of the Janus particle because the rate constants $\kappa_{\pm}$ vanish on the noncatalytic hemisphere.

\subsection{Linearization and induced force density}

The aforementioned stochastic partial differential equations are nonlinear and they are linearized in order to obtain their solutions.\cite{MB74,BM74}  The linearization is justified if the Reynolds number,
\be
{\rm Re}\equiv \frac{V_{\rm d}R}{\nu} \ll 1 \, ,
\label{Reynolds}
\ee
so that the flow is laminar. Here $V_{\rm d}$ is the diffusiophoretic velocity of the Janus particle and $\nu=\eta/\rho$ is the kinematic viscosity.  The condition~(\ref{Reynolds}) is well satisfied for micron-size Janus motors with typical velocities $V_{\rm d}\sim 10^{-5}\, {\rm m/s}$ in water solutions where $\nu\simeq 10^{-6}\, {\rm m}^2/{\rm s}$, since ${\rm Re}\sim 10^{-5}$.  In this low Reynolds number regime the advective term ${\bf v}\cdot\pmb{\nabla}$ is negligible in comparison with the time derivative $\partial_t$, which is of order $\nu/R^2$.

Following Bedeaux and Mazur,\cite{BM74} the problem is reformulated by introducing an induced force density field ${\bf f}_{\rm ind}({\bf r},t)$, so that the linearized equation of motion of the fluid becomes
\be
\rho\, \partial_t{\bf v}({\bf r},t) = -{\rm div}\, {\boldsymbol{\mathsf P}}({\bf r},t)+{\bf f}_{\rm ind}({\bf r},t), \label{lin-NS}
\ee
for an incompressible fluid where Eq.~(\ref{div.v=0}) holds, and with the pressure tensor given in Eq.~(\ref{Pij}).  The induced force density field is chosen to vanish in the fluid, ${\bf f}_{\rm ind}({\bf r},t)=0$ if $r>R$, and to comply with the constraints that the velocity field satisfies the linear equation ${\bf v}({\bf r},t) = {\bf V}(t) + \pmb{\Omega}(t)\times {\bf r}$ if $r\leq R$, and the hydrostatic pressure is zero in the solid particle, $P({\bf r},t)=0$ if $r<R$. As a consequence of Eq.~(\ref{lin-NS}), the induced force density in the solid particle is given by
\be
{\bf f}_{\rm ind}({\bf r},t) = \rho \, \frac{d}{dt} \left[ {\bf V}(t) + \pmb{\Omega}(t)\times {\bf r}\right] \quad\mbox{if}\quad r<R \, ,
\ee
but it is singular on the interface.\cite{MB74}

The linearized problem is solved by Fourier transformation in time,
\be
{\bf v}({\bf r},\omega) = \int_{-\infty}^{+\infty} dt \, {\rm e}^{i\omega t} \, {\bf v}({\bf r},t),
\ee
for the velocity and other fields so that Eq.~(\ref{lin-NS}), using Eqs.~(\ref{div.v=0}) and  (\ref{Pij}), becomes
\bea
&& \left(-i\omega\rho-\eta\nabla^2\right) {\bf v}({\bf r},\omega) = -\pmb{\nabla}P({\bf r},\omega)  \nonumber\\
&&\qquad\qquad\qquad\qquad\qquad-\pmb{\nabla}\cdot\pmb{\pi}({\bf r},\omega)+{\bf f}_{\rm ind}({\bf r},\omega) , \label{lin-NS-f}\\
&& \nabla^2P({\bf r},\omega)  = -\pmb{\nabla}\pmb{\nabla}:\pmb{\pi}({\bf r},\omega)+\pmb{\nabla}\cdot {\bf f}_{\rm ind}({\bf r},\omega) . \label{div.v-f}
\eea
Introducing the Green function,
\be
G({\bf r},\omega) = \frac{1}{4\pi\eta r} \, \exp{(-\alpha r)} \quad \mbox{with} \quad \alpha = \sqrt{-i\omega/\nu}  ,
\ee
(${\rm Re}\,\alpha \geq 0$), the solution of Eqs.~(\ref{lin-NS-f}) and~(\ref{div.v-f}) can be expressed as
\bea
&&{\bf v}({\bf r},\omega) ={\bf v}_0({\bf r},\omega) +\int d{\bf r}' \biggl\{ G({\bf r}-{\bf r}',\omega) \\ &&\quad +\frac{1}{\alpha^2} \, \frac{\partial}{\partial{\bf r}'}\frac{\partial}{\partial{\bf r}'} \left[ G({\bf r}-{\bf r}',0)-G({\bf r}-{\bf r}',\omega)\right]\biggr\}\cdot{\bf f}_{\rm ind}({\bf r}',\omega) \nonumber
\eea
in terms of the induced force density and the fluctuating velocity field ${\bf v}_0$ in the absence of the particle.\cite{MB74,BM74,ABM75}  The correlation functions of the Gaussian white noise velocity field ${\bf v}_0$ may be  determined directly from Eq.~(\ref{pi-correl}).

\subsection{Force of the fluid on the Janus particle}

The fluctuating force exerted by the fluid on the Janus particle can be expressed in terms of the induced force density as~\cite{BM74}
\bea
{\bf F}(\omega) &=& -\int_{\cal S} {\boldsymbol{\mathsf P}}({\bf r},\omega)\cdot{\bf n} \, dS
= - \int_{r\leq R} d{\bf r} \, \pmb{\nabla} \cdot {\boldsymbol{\mathsf P}}({\bf r},\omega)\nonumber\\
&=& -i\omega\rho\, \frac{4\pi R^3}{3}\, {\bf V}(\omega) - \int_{r\leq R} d{\bf r} \, {\bf f}_{\rm ind}({\bf r},\omega).
\label{force-f}
\eea
The relation between the induced force density and the unperturbed velocity field~${\bf v}_0$ remains to be determined.  For this purpose, the Fourier transforms of the boundary conditions~(\ref{v-bc2}) and~(\ref{v-bc1}) are averaged over the surface
to obtain
\bea
&& \overline{{\boldsymbol{\mathsf 1}}_{\bot}\cdot{\bf v}({\bf r},\omega)}^{\rm s}-b\, \overline{{\boldsymbol{\mathsf 1}}_{\bot}\cdot\left[\pmb{\nabla}{\bf v}({\bf r},\omega) +\pmb{\nabla}{\bf v}({\bf r},\omega)^{\rm T}\right]\cdot{\bf n}}^{\rm s}\\
&& +\, \overline{{\boldsymbol{\mathsf 1}}_{\bot}\cdot{\bf v}^{\rm s}_{\rm fl}({\bf r},\omega)}^{\rm s}+\sum_k  b_k \, \overline{{\boldsymbol{\mathsf 1}}_{\bot}\cdot\pmb{\nabla}c_k({\bf r},\omega)}^{\rm s} =  \frac{2}{3}\, {\bf V}(\omega) \, , \nonumber
\label{v-bc2-f} \\
&& \overline{{\bf n}{\bf n}\cdot{\bf v}({\bf r},\omega)}^{\rm s} = \frac{1}{3}\, {\bf V}(\omega) \, , \label{v-bc1-f}
\eea
where
\be
\overline{(\cdot)}^{\rm s} = \frac{1}{4\pi R^2} \int_{r=R} (\cdot ) \, dS\, .
\ee
In writing these equations it has been assumed that the diffusiophoretic constants $b_k$ take uniform values on the entire spherical surface of the particle, otherwise they should be included in the surface average on the left side of Eq.~(\ref{v-bc2-f}). Using the identity $R\int_{r=R} {\bf n}{\bf n}\cdot{\bf v}\,dS = \int_{r\leq R}{\bf v}\, d{\bf r}$, Eq.~(\ref{v-bc1-f}) can be written as the volume average of the velocity field:
\be
{\bf V}(\omega)=\overline{{\bf v}({\bf r},\omega)}^{\rm v} = \frac{3}{4\pi R^3} \int_{r\leq R} {\bf v}({\bf r},\omega) \, d{\bf r} \, .
\ee

From this point the calculations in Sec.~3 of Ref.~\onlinecite{ABM75} can be followed step-by-step to obtain the force~(\ref{force-f}) in the form of a generalized Fax\'en theorem that includes contributions from diffusiophoresis. This force can be written as the sum of three terms,
\begin{equation}\label{force-f-2}
{\bf F}(\omega)= -\gamma(\omega){\bf V}(\omega) +{\bf F}_{\rm d}(\omega) +{\bf F}_{\rm fl}(\omega).
\end{equation}
The first term is the frequency-dependent Stokes drag force $-\gamma(\omega){\bf V}(\omega)$ where $\gamma(\omega)$ is the frequency-dependent Stokes friction coefficient,
\begin{equation}
\gamma(\omega)=6 \pi \eta R \biggl[\frac{(1+\alpha R)(1+2b/R)}{1+b(3+\alpha R)/R} +\frac{\alpha^2 R^2}{9}  \biggr].
\label{g(w)}
\end{equation}
The second term is the frequency-dependent diffusiophoretic force,
\begin{equation}\label{force-d-1}
{\bf F}_{\rm d}(\omega)=\frac{6 \pi \eta R\, (1+\alpha R)}{1+b(3+\alpha R)/R}\, \sum_k  b_k \, \overline{{\boldsymbol{\mathsf 1}}_{\bot}\cdot\pmb{\nabla}c_k({\bf r},\omega)}^{\rm s},
\end{equation}
while the third term is the Langevin fluctuating force,
\begin{eqnarray}
&&{\bf F}_{\rm fl}(\omega)= 6\pi\eta R \biggl\{\frac{\alpha^2R^2}{3}\, \overline{{\bf v}_0({\bf r},\omega)}^{\rm v} \\
&&\quad + \frac{1+\alpha R}{1+b(3+\alpha R)/R} \Big[1 + \frac{b}{R} \Big(2-R\frac{\partial}{\partial R}\Big)\Big] \overline{{\bf v}_0({\bf r},\omega)}^{\rm s}\nonumber\\
&& \qquad + \frac{1+\alpha R}{1+b(3+\alpha R)/R}\, \overline{{\boldsymbol{\mathsf 1}}_{\bot}\cdot{\bf v}^{\rm s}_{\rm fl}({\bf r},\omega)}^{\rm s}\biggr\} , \nonumber
\end{eqnarray}
which depends on the fluctuating unperturbed velocity field ${\bf v}_0({\bf r},\omega)$ and can be shown to obey the fluctuation-dissipation theorem,\cite{BM74,BAM77}
\bea
&&\langle {\bf F}_{\rm fl}(\omega)\rangle = 0 \qquad\mbox{and} \\
&&\langle {\bf F}_{\rm fl}(\omega)\, {\bf F}_{\rm fl}^*(\omega')\rangle = 4\pi\, k_{\rm B}T \, {\rm Re}\, \gamma(\omega) \, \delta(\omega-\omega') \, {\boldsymbol{\mathsf 1}} \, .\nonumber
\eea
The expressions for the Stokes drag force and Langevin fluctuating force given above were obtained earlier.\cite{ABM75,BAM77}  The frequency-dependent translational friction coefficient for stick boundary conditions is known,\cite{LL87} and agrees with Eq.~(\ref{g(w)}) for $b=0$.

The equation of motion for a Brownian particle corresponding to the frequency-dependent friction has a memory kernel with a long-time tail and an extra acceleration term giving an effective mass $m+m_{\rm fluid}/2$ where $m_{\rm fluid}$ is the mass of the displaced fluid.\cite{ZB70,HM73,D74,LR13}  The long-time tail also manifests itself in the time-dependent correlation function of the fluctuating force.  In the noiseless limit, the equation of motion is consistent with that in Refs.~\onlinecite{CL56,MR83}.  The effects of the long-time tail become negligible at low frequency if the Lorentz condition $m\gg m_{\rm fluid}$ is satisfied, in which case the standard Langevin equation is recovered for a Brownian particle without the diffusiophoretic force.  A crossover to the Langevin low-frequency regime occurs for $\vert\alpha R\vert\sim 1$ around the frequency $\omega\sim\nu/R^2$ characteristic of shear viscosity.  This low-frequency limit will be considered in Sec.~\ref{Low-freq}.

If the compressibility of the fluid is taken into account, modifications appear beyond the frequency
$\omega\simeq v_{\rm sound}/R$ where $v_{\rm sound}$ is the sound velocity.\cite{BM74c} This is a very large frequency in the case of water where $\omega\simeq 1.5\times 10^{9}\, {\rm s}^{-1}$, so that these effects will be neglected here.

\subsection{Torque of the fluid on the particle}

A Fax\'en-like theorem for the torque on a spherical particle using methods similar to those for the force was derived earlier.\cite{H75,F76a,F76b} Extending such calculations to include the diffusiophoretic contribution we find that the torque can also be written as the sum of three contributions,
\begin{equation} \label{torque-f}
{\bf T}(\omega) = -\gamma_{\rm r}(\omega) \, \pmb{\Omega}(\omega)  + {\bf T}_{{\rm d}}(\omega)+{\bf T}_{\rm fl}(\omega),
\end{equation}
with the frequency-dependent rotational friction coefficient given by
\be \label{rot-frict}
\gamma_{\rm r}(\omega) = 8\pi\eta R^3 (1-3\,\xi) \frac{1+\alpha R+(\alpha R)^2/3}{1+\alpha R+\xi (\alpha R)^2} ,
\ee
where $\xi = b/(R+3b)$.\cite{F76b} For stick boundary condition ($b=0$), the known expression\cite{LL87} is recovered.

The diffusiophoretic torque is
\begin{equation}
{\bf T}_{{\rm d}}(\omega)=\frac{3}{2R^2} \, \gamma_{\rm r}(\omega)  \sum_k  b_k  \overline{{\bf r}\times\pmb{\nabla}c_k({\bf r},\omega)}^{\rm s},
\end{equation}
while the random torque ${\bf T}_{\rm fl}(\omega)$ is a Gaussian white noise process with
\bea
&&\langle {\bf T}_{\rm fl}(\omega)\rangle = 0 , \\
&&\langle {\bf T}_{\rm fl}(\omega)\, {\bf T}_{\rm fl}^*(\omega')\rangle = 4\pi\, k_{\rm B}T \, {\rm Re}\, \gamma_{\rm r}(\omega) \, \delta(\omega-\omega') \, {\boldsymbol{\mathsf 1}}  .\nonumber
\eea
The frequency dependence of the torque undergoes a similar crossover as for the force around the viscosity characteristic frequency $\omega\sim\nu/R^2$.  The low-frequency limit will be taken in Sec.~\ref{Low-freq}.

\subsection{Reaction rate and concentration fields}

The advection-diffusion equations~(\ref{diff-eq-1})-(\ref{diff-eq-2}) for the concentration fields and their boundary conditions~(\ref{bc-cA})-(\ref{bc-cB}) can be linearized if the P\'eclet numbers are small enough:
\be
{\rm Pe}_k\equiv \frac{V_{\rm d}R}{D_k} \ll 1 \, .
\ee
This condition holds for a micron-size Janus particle moving at the velocity $V_{\rm d}\sim 10^{-5}\, {\rm m/s}$ where the solute molecular diffusion coefficients are of order $D_k\sim 10^{-9}\, {\rm m}^2/{\rm s}$, so that ${\rm Pe}_k\sim 10^{-2}$.  Under such circumstances the advective term in Eq.~(\ref{diff-eq-2}) is negligible and we obtain the linearized fluctuating diffusion equation:
\be
\partial_t \, c_k =D_k \nabla^2 c_k -\pmb{\nabla}\cdot\pmb{\eta}_k \, .
\label{diff-eq}
\ee

In terms of the Fourier transforms of the concentration fields,
\be
c_k({\bf r},\omega) = \int_{-\infty}^{+\infty} dt \, {\rm e}^{i\omega t} \, c_k({\bf r},t),
\ee
Eq.~(\ref{diff-eq}) can be written as
\be
\left(-i\omega-D_k\nabla^2\right) c_k({\bf r},\omega) +\pmb{\nabla}\cdot\pmb{\eta}_k({\bf r},\omega)=\sigma_k({\bf r},\omega),
\label{lin-diff-eq-f}
\ee
where the source term $\sigma_k$ is defined on the reactive interface and may be determined using the linearized boundary conditions~(\ref{bc-cA}) and~(\ref{bc-cB}). This source plays a role that is analogous to the induced force density.
The diffusion Green functions satisfy the equation
\be
\left(-i\omega-D_k\nabla^2\right) G_k({\bf r},\omega) =\delta({\bf r})
\ee
and are given by
\be
G_k({\bf r},\omega) = \frac{1}{4\pi D_k r} \, \exp{(-\alpha_k r)} \quad \mbox{with} \quad \alpha_k = \sqrt{-i\omega/D_k} ,
\ee
(${\rm Re}\,\alpha_k \geq 0$). The solution of Eq.~(\ref{lin-diff-eq-f}) can be expressed in terms of these Green functions as
\be
c_k({\bf r},\omega) =c_{k0}({\bf r},\omega) +\int d{\bf r}' \, G_k({\bf r}-{\bf r}',\omega) \, \sigma_k({\bf r}',\omega) ,
\label{c-G}
\ee
where $c_{k0}({\bf r},\omega)$ is the unperturbed fluctuating concentration field in the absence of the effects of $\sigma_k$, and is given by the solution of
\be
\left(-i\omega-D_k\nabla^2\right) c_{k0}({\bf r},\omega) +\pmb{\nabla}\cdot\pmb{\eta}_k({\bf r},\omega)=0 \, .
\ee
These equations can be solved by methods that are similar to those described above for the velocity field by considering the averages of Eq.~(\ref{c-G}) over the volume and the surface of the Janus particle.\cite{LK79} The following expression for the frequency-dependent reaction rate~(\ref{W}) is found:
\bea
W(\omega) &=&  4\pi R D_k  \nu_k  (1+\alpha_k R) \, \left[\overline{c_k({\bf r},\omega)}^{\rm s} -\overline{c_{k0}({\bf r},\omega)}^{\rm s}\right] \nonumber\\
&&\qquad\qquad\qquad+i\omega \, \frac{4\pi R^3}{3}\, \nu_k \,  \overline{c_{k0}({\bf r},\omega)}^{\rm v}\, .\label{rate-f}
\eea
Contrary to Eq.~(\ref{force-f-2}) for the force or Eq.~(\ref{torque-f}) for the torque, the expression~(\ref{rate-f}) does not have a closed form. If the catalytic surface was spherical, the frequency-dependent rate would also be given by
\be
W(\omega)=4\pi R^2\Big[ \kappa_+\overline{c_{\rm A}({\bf r},\omega)}^{\rm s}-\kappa_-\overline{c_{\rm B}({\bf r},\omega)}^{\rm s}\Big] ,
\label{W-sph}
\ee
so that Eqs.~(\ref{rate-f}) and~(\ref{W-sph}) could be combined to obtain an expression involving the surface and volume averages of the unperturbed concentration fields, together with the terms of diffusiophoretic origin.  However, Eq.~(\ref{W-sph}) does not hold in the hemispherical geometry of a Janus particle because the rate constants $\kappa_{\pm}$ vary along the surface of the particle.  Therefore, the inversion of Eq.~(\ref{rate-f}) is not straightforward for a Janus particle.  However, we see that $\alpha_k$ plays a role similar to $\alpha$ for the velocity field; thus, the frequency dependence should present a crossover for $\vert\alpha_kR\vert\sim 1$, corresponding to the frequency $\omega\sim D_k/R^2$ characteristic of the diffusion of molecular species $k$.  The low-frequency limit will be analyzed in Sec.~\ref{Low-freq}.

\section{Low-frequency limit}
\label{Low-freq}

In this section, we consider the low-frequency limit for the frequency-dependent force~(\ref{force-f-2}), torque~(\ref{torque-f}), and rate~(\ref{rate-f}).  In this regime, the two conditions $\vert\alpha R\vert \ll 1$ and $\vert\alpha_k R\vert\ll 1$ are satisfied, corresponding to the frequency range where both $\omega\ll \nu/R^2$ and $\omega\ll D_k/R^2$ apply.

\subsection{Translation and rotation}

In the low-frequency limit $\omega\ll \nu/R^2$, the force~(\ref{force-f-2})
is expressed in terms of frequency-independent translational friction coefficient
\be \label{t-friction}
\gamma= 6\pi\eta R \, \frac{1+2b/R}{1+3b/R},
\ee
which is related to the Janus particle diffusion coefficient by the Einstein formula, $D\equiv k_{\rm B}T/\gamma = (\beta \gamma)^{-1}$.  The diffusiophoretic force has the simpler form,
\be
{\bf F}_{\rm d}(\omega) = \frac{6\pi\eta R}{1+3b/R} \sum_k  b_k \, \overline{{\boldsymbol{\mathsf 1}}_{\bot}\cdot\pmb{\nabla}c_k({\bf r},\omega)}^{\rm s} .
\label{Fd}
\ee
Using the expression~(\ref{b_k}) for the diffusiophoretic constants, $b_k$, the diffusiophoretic force and friction coefficient adopt the forms
\be
{\bf F}_{{\rm d}}(\omega) =6\pi R \, k_{\rm B}T \sum_k K^{(1)}_k  \, \overline{{\boldsymbol{\mathsf 1}}_{\bot}\cdot\pmb{\nabla}c_k({\bf r},\omega)}^{\rm s}
\ee
and $\gamma=6 \pi \eta R$ for stick boundary conditions, $b=0$, and
\be
{\bf F}_{{\rm d}}(\omega) =2\pi R^2  k_{\rm B}T \sum_k K^{(0)}_k \, \overline{{\boldsymbol{\mathsf 1}}_{\bot}\cdot\pmb{\nabla}c_k({\bf r},\omega)}^{\rm s}
\ee
and $\gamma=4 \pi \eta R$ for perfect slip boundary conditions, $b=\infty$. The diffusiophoretic force is well defined in both limits.

Also, in the low-frequency limit, the torque~(\ref{torque-f}) is given in terms of the frequency-independent rotational friction coefficient
\be \label{r-friction}
\gamma_{\rm r}=\frac{8\pi\eta  R^3}{1+3b/R},
\ee
and the diffusiophoretic torque is
\be \label{rot-diff-t}
{\bf T}_{{\rm d}}(\omega)= \frac{12\pi\eta  R}{1+3b/R} \, \sum_k  b_k \, \overline{{\bf r}\times\pmb{\nabla}c_k({\bf r},\omega)}^{\rm s}.
\ee
In the limit of perfect stick we have
\be
{\bf T}_{{\rm d}}(\omega)=12\pi R \, k_{\rm B}T  \, \sum_k K^{(1)}_k \, \overline{{\bf r}\times\pmb{\nabla}c_k({\bf r},\omega)}^{\rm s}
\ee
and $\gamma_{\rm r}=8\pi\eta R^3$, which is consistent with a result obtained previously in this limit.\cite{AP91} For perfect slip we have
\be
{\bf T}_{{\rm d}}(\omega) = 4\pi R^2 k_{\rm B}T  \, \sum_k K^{(0)}_k \, \overline{{\bf r}\times\pmb{\nabla}c_k({\bf r},\omega)}^{\rm s}
\ee
and $\gamma_{\rm r}=0$. For a spherical Janus particle, the diffusiophoretic torque vanishes by cylindrical symmetry, ${\bf T}_{\rm d}(\omega)=0$, so that only the frictional torque due to viscosity remains.

We note that, in the low-frequency domain $\omega \ll \nu/R^2$, long-time tail effects play a negligible role.  For a micrometric particle with $R=10^{-6}\, {\rm m}$ in water, this range extends up to $\omega\ll 10^6\, {\rm s}^{-1}$, corresponding to the microsecond time scale.

\subsection{Diffusion and reaction}
\label{sph-J-part}

For diffusion, the low-frequency regime where $\vert\alpha_k R\vert\ll 1$ for all the species $k$ corresponds to the range $\omega\ll D_k/R^2\sim 10^3\, {\rm s}^{-1}$ for molecular diffusivities of the order of $D_k\sim 10^{-9}\, {\rm m}^2/{\rm s}$.  In this regime, the concentration fields take their static profile around the Janus particle and can thus be determined by standard methods.\cite{HSK16,OPD17}

For a spherical Janus particle of radius $R$, the stationary concentrations $c_k(r,\theta)$ $(k={\rm A},\, {\rm B})$ may be obtained by solving the diffusion equations, $\nabla^2 c_k = 0$, subject to the boundary conditions,
\bea
&& D_k \partial_r c_k\vert_{r=R} = -\nu_k\, \chi(\theta) \, \left(\kappa_+ c_{\rm A}-\kappa_- c_{\rm B}\right)_{r=R}  , \nonumber\\
&& c_k\vert_{r=\infty} = \bar{c}_k \, ,
\eea
with $\chi(\theta) = H(\cos\theta)$ where the Heaviside function $H(\xi)$ takes the values $H(\xi)=1$ on the catalytic hemisphere and $H(\xi)=0$ on the chemically inactive hemisphere.  Here, we discard the diffusiophoretic terms in the boundary conditions in order to obtain an analytical expression for the main contribution to the reaction rate.  The diffusiophoretic effect on the reaction rate will be restored in Sec.~\ref{Langevin} by using the Onsager-Casimir symmetry, but at the level of the overdamped Langevin equations.

The concentration fields of species $k$ around a spherical Janus particle of radius $R$ are given by
\be \label{cA-f}
c_k(r,\theta) =  \bar{c}_k +\nu_k\frac{R}{D_k} \Big(\kappa_+\bar{c}_{\rm A}-\kappa_-\bar{c}_{\rm B}\Big) f(r,\theta) \, ,
\ee
where the function $f(r,\theta)$ satisfies the diffusion equation $\nabla^2f=0$ with the boundary conditions $R\,\partial_rf\vert_R = H(\cos\theta)\left( {\rm Da}\, f-1\right)_R$ and $f\vert_{\infty}=0$ in spherical coordinates $(r,\theta,\phi)$ aligned parallel to the particle axis.  The boundary condition at the particle surface $r=R$ involves the dimensionless Damkh\"oler number
\be
{\rm Da} \equiv R\left(\frac{\kappa_+}{D_{\rm A}} + \frac{\kappa_-}{D_{\rm B}}\right) .
\label{Da}
\ee
Letting $k_\pm=4 \pi R^2 \kappa_\pm$, we see that this number takes the form ${\rm Da}=k_+/k_{D_{\rm A}}+k_-/k_{D_{\rm B}}$ where $k_{D_k}= 4 \pi D_k R$ are the Smoluchowski diffusion-controlled rate coefficients. In the reaction-limited regime $k_\pm \ll k_{D_k}$ and ${\rm Da}\ll 1$, while in the diffusion-controlled regime $k_\pm \gg k_{D_k}$ and ${\rm Da}\gg 1$.

To obtain the solution, $f(r,\theta)$ is expanded in Legendre functions $P_l(\xi)$, with $\xi=\cos \theta$,
\be
f(r,\theta) =\sum_{l=0}^{\infty} a_l \, P_l(\xi) (R/r)^{l+1}.
\label{f}
\ee
The coefficients are given by $a_l = \sum_{l'=0}^{\infty} \left({\boldsymbol{\mathsf M}}^{-1}\right)_{ll'} \, {\mathcal A}_{l'}$,
where $M_{ll'} = 2(2l+1)^{-1} \left( l+1 \right) \delta_{ll'} + {\rm Da}\, {\mathcal B}_{l l'}$
with ${\mathcal A}_l=\int_0^1 d\xi \, P_l(\xi)$ and ${\mathcal B}_{l l'}=\int_0^1 d\xi \, P_l(\xi) \, P_{l'}(\xi)$.

Using these results the mean value of the reaction rate~(\ref{W}) may be written as
\be
W_{\rm rxn} = \Gamma \left(\kappa_+\bar{c}_{\rm A} -  \kappa_-\bar{c}_{\rm B}\right) ,
\label{Wrxn}
\ee
with $\Gamma=2\pi R^2(1-{\rm Da}\,\gamma_J)$ where
\be
\gamma_J = \frac{1}{2\pi R^2 } \int_{r=R} dS \, f(r,\theta)\, H(\cos\theta)= \sum_{l=0}^{\infty} a_l \, {\mathcal A}_l\, .
\label{gamma_J}
\ee
Equivalently, the mean reaction rate is given by Eq.~(\ref{rate-f}) in the limit $\omega=0$, showing that $\Gamma=4\pi R^2 a_0$.  The surface average of the aforementioned boundary condition satisfied by $f(r,\theta)$ at $r=R$ confirms that $a_0=(1-{\rm Da}\, \gamma_J)/2$.  The dimensionless coefficients $a_0$ and $\gamma_J$ may be obtained by numerical evaluations as a function of the Damkh\"oler number (\ref{Da}). Appendix~\ref{AppA} gives additional details pertaining to the dependence of $\gamma_J$ on ${\rm Da}$.

At thermodynamic equilibrium, the concentrations $\bar{c}_{k}$ satisfy the Guldberg-Waage condition:
\be
\frac{\bar{c}_{\rm A, eq}}{ \bar{c}_{\rm B, eq}} = \frac{\kappa_-}{\kappa_+} = \exp\frac{\Delta\mu^0}{k_{\rm B}T} \, ,
\label{GW-eq}
\ee
where $\Delta\mu^0=\mu^0_{\rm B}-\mu^0_{\rm A}$ is the standard free energy of the reaction ${\rm A}\to{\rm B}$.  The free energy of the reaction is related to the concentrations by $\Delta\mu=\Delta\mu^0 + k_{\rm B}T \, \ln(\bar{c}_{\rm B}/\bar{c}_{\rm A})$, which vanishes at equilibrium, $\Delta\mu_{\rm eq}=0$, because of Eq.~(\ref{GW-eq}). The dimensionless affinity driving the reaction out of equilibrium is defined in general as
\be
A_{\rm rxn} \equiv \ln \frac{\kappa_+\bar{c}_{\rm A}}{\kappa_-\bar{c}_{\rm B}} = -\frac{\Delta\mu}{k_{\rm B}T} \, ,
\ee
which is positive (resp. negative) for the reaction running in the direction ${\rm A}\to{\rm B}$ (resp. ${\rm B}\to{\rm A}$), and vanishes at equilibrium.

In the following, we consider the reaction in the linear regime close to equilibrium where the deviations of the concentrations from their equilibrium values, $\delta \bar{c}_k \equiv \bar{c}_k-\bar{c}_{k,{\rm eq}}$ are small: $\vert\delta \bar{c}_k\vert \ll \bar{c}_{k,{\rm eq}}$.  In this regime, the chemical affinity can be approximated as
\be
A_{\rm rxn} \simeq \frac{\delta\bar{c}_{\rm A}}{\bar{c}_{\rm A, eq}} - \frac{\delta\bar{c}_{\rm B}}{\bar{c}_{\rm B, eq}} \, ,
\label{Arxn-lin}
\ee
up to terms of second order in the concentration deviations $\delta \bar{c}_k$.

Introducing the reaction diffusivity
\be \label{D-rxn-2}
D_{\rm rxn} \equiv \frac{\Gamma}{2} \left(\kappa_+\bar{c}_{\rm A} + \kappa_-\bar{c}_{\rm B}\right) \, ,
\ee
associated with the reaction rate~(\ref{Wrxn}), the chemical affinity~(\ref{Arxn-lin}) may also be written close to equilibrium as
\be
A_{\rm rxn} = \frac{W_{\rm rxn}}{D_{\rm rxn}} \, ,
\label{Arxn}
\ee
up to terms with higher powers in the reaction rate~(\ref{Wrxn}).

\subsection{The diffusiophoretic force}

Since the concentration fields in Eq.~(\ref{cA-f}) are now known, we may write a more explicit expression for the diffusiophoretic force~(\ref{Fd}):
\be \label{diff-force}
{\bf F}_{\rm d} = \frac{4\pi\eta R}{1+3b/R} \left(\frac{ b_{\rm B}}{D_{\rm B}}-\frac{ b_{\rm A}}{D_{\rm A}}\right)\left(\kappa_+\bar{c}_{\rm A} -  \kappa_-\bar{c}_{\rm B}\right)  a_1 {\bf u},
\ee
where we used the fact that
\be
\overline{{\boldsymbol{\mathsf 1}}_{\bot}\cdot\pmb{\nabla}f(r,\theta)}^{\rm s} = \frac{{\bf u}}{R} \int_{-1}^{+1} d\xi \, \xi f(R,\theta) =\frac{2}{3} \frac{a_1}{R}{\bf u}\, .
\label{int-u}
\ee
The main features of the dependence of the coefficient $a_1$ on $\rm Da$ are given in Appendix~\ref{AppA}.
From Eq.~(\ref{diff-force}), we see that the diffusiophoretic force is aligned parallel to the particle axis
\be \label{Fd-Wrxn}
{\bf F}_{\rm d} = F_{\rm d} \, {\bf u} \equiv \gamma\, \chi \, W_{\rm rxn} \, {\bf u} \, ,
\ee
where we have rewritten $F_{\rm d}$ in the second equality to define the diffusiophoretic parameter $\chi$,
\be
\chi \equiv  \frac{F_{\rm d}}{\gamma \, W_{\rm rxn}}
= \frac{a_1}{a_0 6\pi R^2 \left(1+2b/R\right)} \, \left(\frac{ b_{\rm B}}{D_{\rm B}}-\frac{ b_{\rm A}}{D_{\rm A}}\right) ,
\label{chi}
\ee
which will be useful in what follows. Using the approximation given by Eq.~(\ref{b_k}) for the diffusiophoretic constants, we have that
\bea
\chi &\simeq& \frac{a_1 k_{\rm B}T}{a_0 6\pi \eta R^2 \left(1+2b/R\right)} \nonumber\\
&&\qquad\times \left(\frac{K_{\rm B}^{(1)} + b K_{\rm B}^{(0)}}{D_{\rm B}}-\frac{K_{\rm A}^{(1)} + b K_{\rm A}^{(0)}}{D_{\rm A}}\right) .
\eea
We see that $F_{\rm d}$ remains finite in the limits of perfect stick ($b=0$) and perfect slip ($b\to\infty$) boundary conditions.

Our calculation does not include corrections due to non-vanishing P\'eclet numbers.\cite{AP91,AB06}

\section{Coupled Langevin equations in the overdamped regime}
\label{Langevin}

In an earlier study overdamped Langevin equations for translation, rotation and reaction were written and used to derive a nonequilibrium mechanochemical fluctuation theorem for diffusiophoretic Janus motors.\cite{GK17} In this section, we show how these Langevin equations can be derived from the fluctuating thermodynamics formalism presented above, and  deduce further results about the nonequilibrium dynamics of Janus motors.

\subsection{Translation}

The overdamped regime, where the friction due to viscosity dominates the inertial effects, lies in the domain where $\omega\ll 6\pi \eta R/m =4.5(\rho/\rho_{\rm sol})(\nu/R^2)$.  This corresponds to the condition $\omega\ll 10^6\, {\rm s}^{-1}$ for a solid particle with mass density $\rho_{\rm sol}\simeq 4.5\, \rho$.  In this case, the overdamped regime essentially coincides with the low-frequency domain where the Lorentz condition is satisfied and the long-time tail effects can be neglected.

Using the low-frequency limit of the force, we may immediately write the Langevin equation for a spherical Janus particle of mass $m$ as
\be
m\frac{d{\bf V}}{dt} = -\gamma\, {\bf V}  + {\bf F}_{\rm d} + {\bf F}_{\rm ext} + {\bf F}_{\rm fl}(t) \, ,
\label{Langevin-eq-transl}
\ee
where the translational friction coefficient $\gamma$ is given in Eq.~(\ref{t-friction}) and the diffusiophoretic force in Eq.~(\ref{Fd-Wrxn}).  We have also added an external force, ${\bf F}_{\rm ext}$. We notice that, in general, the Langevin equation will contain an additional term that accounts for coupling between translation and rotation, $- \pmb{\gamma}_{\rm t,r}\cdot\pmb{\Omega}$, where $\pmb{\gamma}_{\rm t,r}$ is the translation-rotation friction tensor. Since we consider a spherical Janus particle we have $\pmb{\gamma}_{\rm t,r}=0$.

According to the Langevin equation, the particle velocity becomes Maxwellian over the thermalization time scale $\tau_{\rm t} = m/\gamma$.  Since the mass is given by $m = (4\pi/3)R^3  \rho_{\rm sol}$ in terms of the particle mass density $\rho_{\rm sol}$, the thermalization time becomes $\tau_{\rm t} =2\rho_{\rm sol}R^2/(9\eta)$ if $b\ll R$.  For a silica micrometric particle in water at $20^{\circ}$C, the parameters take the values $\rho_{\rm sol}\simeq 1522\, {\rm kg}/{\rm m}^3$, $\eta\simeq 10^{-3}\, {\rm N\, s}/{\rm m}^2$, and $R=10^{-6}\, {\rm m}$, so that the thermalization time is estimated to be $\tau_{\rm t} \simeq 3\times 10^{-7}\, {\rm s}$.  This time scale is short enough to justify considering the overdamped limit where the velocity distribution remains Maxwellian.

In the overdamped limit, the inertial term proportional to the mass $m$ is negligible, so that the Langevin equation for a spherical Janus particle reduces to
\be
\frac{d{\bf r}}{dt} = {\bf V}_{\rm d}  + \beta D \, {\bf F}_{\rm ext} + {\bf V}_{\rm fl}(t) \, ,
\label{eq-r}
\ee
where we used $\gamma^{-1}=\beta D$ to write the coefficient of ${\bf F}_{\rm ext}$. The fluctuating velocity field is ${\bf V}_{\rm fl}(t) ={\bf F}_{\rm fl}(t)/\gamma$, and satisfies
\bea
&&\langle {\bf V}_{\rm fl}(t)\rangle = 0 \ ,  \nonumber\\
&&\langle {\bf V}_{\rm fl}(t) {\bf V}_{\rm fl}(t')\rangle = 2D \, \delta(t-t')\, {\boldsymbol{\mathsf 1}}.
\eea
The diffusiophoretic velocity, ${\bf V}_{\rm d} ={\bf F}_{\rm d}/\gamma$, is
\be \label{Vd}
{\bf V}_{\rm d} =\frac{1}{1+2b/R} \sum_k  b_k \, \overline{{\boldsymbol{\mathsf 1}}_{\bot}\cdot\pmb{\nabla}c_k({\bf r})}^{\rm s} =  V_{\rm d} \, {\bf u},
\ee
where we have used Eq.~(\ref{Fd}). The unit vector ${\bf u} = (\sin\theta \, \cos\phi,\sin\theta \, \sin\phi,\cos\theta)$ is taken to lie along the axis of the Janus particle and oriented from the inert towards the catalytic hemisphere.  The overdamped Langevin equation~(\ref{eq-r}) has standard form used in other studies of active Janus particle dynamics; however, the general expression for the diffusiophoretic velocity~(\ref{Vd}) that includes partial slip differs because of the form of the diffusiophoretic force derived above. In the limit of perfect stick we obtain the usual expression,\cite{A89}
\be \label{Vd-stick}
{\bf V}_{\rm d} = \frac{k_{\rm B}T}{\eta}\sum_k  K_k^{(1)} \, \overline{{\boldsymbol{\mathsf 1}}_{\bot}\cdot\pmb{\nabla}c_k({\bf r})}^{\rm s} ,
\ee
while for perfect slip the result is
\be \label{Vd-slip}
{\bf V}_{\rm d} = \frac{k_{\rm B}T}{\eta}\frac{R}{2} \sum_k  K_k^{(0)} \, \overline{{\boldsymbol{\mathsf 1}}_{\bot}\cdot\pmb{\nabla}c_k({\bf r})}^{\rm s} .
\ee
The expression~(\ref{Vd}) for the diffusiophoretic velocity differs from that obtained in Ref.~\onlinecite{AB06} where the denominator $1+2b/R$ is absent, but appears in our calculation which is based on the generalized Faxen theorem.\cite{MB74,BM74}

\subsection{Rotation}
For a spherical particle with inertia moment $I=2mR^2/5=8\pi\rho_{\rm sol} R^5/15$, the overdamped regime, where the rotational friction due to viscosity dominates the inertial effects, corresponds to the domain where $\omega\ll 8\pi \eta R^3/I =15(\rho/\rho_{\rm sol})(\nu/R^2)\sim 3\times 10^6\, {\rm s}^{-1}$ for a solid particle with mass density $\rho_{\rm sol}\simeq 4.5\, \rho$.  In this case, the overdamped regime also essentially coincides to the low-frequency domain $\omega\ll \nu/R^2$.

Using Eq.~(\ref{torque-f}), in this regime the Langevin equation for rotational motion of a Janus particle is given by
\be
{\boldsymbol{\mathsf I}}\cdot\frac{d\pmb{\Omega}}{dt} = - \gamma_{\rm r}\,\pmb{\Omega} + {\bf T}_{\rm d} + {\bf T}_{\rm ext} + {\bf T}_{\rm fl}(t) \, ,
\label{Langevin-eq-rot}
\ee
where ${\bf T}_{\rm d}$ is a torque due to diffusiophoresis, ${\bf T}_{\rm ext}$ an external torque, and ${\bf T}_{\rm fl}(t)$ the Langevin fluctuating torque.  Similar to the translational Langevin equation, we have omitted a term that couples rotation to translation ($-\pmb{\gamma}_{\rm r,t} \cdot {\bf V}$) since this term vanishes for our Janus particle, as does the diffusiophoretic torque, ${\bf T}_{\rm d}=0$. The rotational velocity determines the time evolution of the unit vector $\bf u$ according to
\be
\frac{d{\bf u}}{dt} = \pmb{\Omega}\times{\bf u} \, .
\ee
If the Janus particle carries a magnetic moment $\pmb{\mu}=\mu{\bf u}$ and the system is subjected to a uniform external magnetic field $\bf B$, the external torque is given by ${\bf T}_{\rm ext} = \mu\, {\bf u}\times{\bf B}$, which has the effect of orienting the magnetic moment (hence the unit vector) in the direction of the magnetic field~$\bf B$.

If $b\ll R$, the thermalization time of the rotational velocity is given by $\tau_{\rm r} = I/\gamma_{\rm r} = \rho_{\rm sol}R^2/(15\eta)  = 3\tau_{\rm t}/10$, which takes the value $\tau_{\rm r} \simeq 10^{-7}\, {\rm s}$ for a micrometric particle in water.  Again, the time scale is short enough to justify taking the overdamped limit:
\bea
\frac{d{\bf u}}{dt} &=& -\frac{1}{\gamma_{\rm r}} \, {\bf u}\times \left[{\bf T}_{\rm ext} + {\bf T}_{\rm fl}(t) \right] \nonumber\\
&=& \frac{\mu}{\gamma_{\rm r}} \left( {\bf B} - {\bf B}\cdot{\bf u}\, {\bf u}\right)-\frac{1}{\gamma_{\rm r}} \, {\bf u}\times  {\bf T}_{\rm fl}(t) \, .
\label{eq-rot}
\eea
Note that this equation does not depend on the particle position or the reactive state so that this stochastic equation is decoupled from the other equations; thus, it drives the direction $\bf u$ independently of what happens for translation and reaction because the Janus particle is spherical.

\subsection{Reaction}

The number $n$ of reactive events during the time interval $[0,t]$ since the beginning of observation is also ruled by a stochastic differential equation:
\be
\frac{dn}{dt} = W_{\rm rxn} + W_{\rm d} + W_{\rm fl}(t) \, ,
\label{eq-n}
\ee
where $W_{\rm rxn}$ is the aforementioned mean reaction rate, $W_{\rm d}$ a contribution from diffusiophoresis to be determined, and $W_{\rm fl}(t)$ a fluctuating rate.  The rate~(\ref{eq-n}) can be written as $dn/dt= -dN_{\rm A}/dt= dN_{\rm B}/dt$ in terms of the numbers of molecules A and B in the solution.

In the overdamped limit, we notice that the contribution $W_{\rm d}$ is reciprocal to the diffusiophoretic term ${\bf V}_{{\rm d}}={\bf F}_{{\rm d}}/\gamma=\chi \, W_{\rm rxn} \, {\bf u}$ in Eq.~(\ref{eq-r}).  Since the variables $\bf r$ and $n$ are even under time reversal, we can use the Onsager symmetry principle in order to determine $W_{\rm d}$.  First, we identify the affinities or generalized thermodynamic forces as the mechanical and chemical affinities. The chemical affinity $A_{\rm rxn}$ was defined earlier in Eq.~(\ref{Arxn}), while the mechanical affinity is ${\bf A}_{\rm mech} = \beta \, {\bf F}_{\rm ext}$.
Gathering the variables and the affinities in the four-dimensional vectors ${\bf X} = ({\bf r} , \, n )$ and ${\bf A} = ({\bf A}_{\rm mech} , \, A_{\rm rxn}  )$, the coupled stochastic equations~(\ref{eq-r}) and~(\ref{eq-n}) can be expressed as $dX_{\alpha}/dt=J_\alpha$ in terms of the currents~(\ref{J-alpha}) in the overdamped regime. Since we know that the position~$\bf r$ is ruled by Eq.~(\ref{eq-r}) and that the matrix of linear response coefficients should be symmetric in order to satisfy Onsager's reciprocal relations, we deduce that
\be
{\boldsymbol{\mathsf L}} = (L_{\alpha\beta}) =
\left(
\begin{array}{cc}
D \, {\boldsymbol{\mathsf 1}} & \chi\, D_{\rm rxn} \, {\bf u} \\
\chi\, D_{\rm rxn} \, {\bf u} & D_{\rm rxn}
\end{array}
\right).
\label{L}
\ee
This matrix is non-negative to satisfy the second law of thermodynamics. As a consequence of Eqs.~(\ref{J-alpha}) and~(\ref{L}), the coupled stochastic differential equations for the position $\bf r$ and the number $n$ are given by
\bea
\frac{d{\bf r}}{dt} &=& \chi \, W_{\rm rxn} \, {\bf u} + \beta D\, {\bf F}_{\rm ext} + {\bf V}_{\rm fl}(t) \, , \label{eq-r-fin}\\
\frac{dn}{dt} &=& W_{\rm rxn} + \beta\chi D_{\rm rxn} {\bf u}\cdot{\bf F}_{\rm ext} + W_{\rm fl}(t) \, , \label{eq-n-fin}
\eea
with the fluctuating velocity ${\bf V}_{\rm fl}(t)$ and the fluctuating reaction rate $W_{\rm fl}(t)$ given by the coupled Gaussian white noise processes characterized by Eq.~(\ref{Gaussian}) with $\left[\delta J_{\alpha}(t)\right]=\left[{\bf V}_{\rm fl}(t),W_{\rm fl}(t)\right]$.  As required, Eq.~(\ref{eq-r-fin}) is identical to Eq.~(\ref{eq-r}) since $V_{\rm d}=\chi W_{\rm rxn}$.  We emphasize that Eqs.~(\ref{eq-r-fin}) and~(\ref{eq-n-fin}) are coupled to Eq.~(\ref{eq-rot}) for rotation. The implication of the reciprocal effect is that an external force combined with an external torque aligning the Janus particle in a preferential orientation can influence the reaction rate.\cite{GK17}

\subsection{The Fokker-Planck equation}

The Fokker-Planck equation governing the time evolution of the probability density $p({\bf r},n,{\bf u};t)$ can be written as
\be
\partial_t p =-\partial_{\bf X}\cdot{\pmb{\cal J}} +\hat{L}_{\rm r} p
\label{FP-eq}
\ee
with the associated current density
\be
\pmb{\cal J} = {\boldsymbol{\mathsf L}}\cdot{\bf A}\, p - {\boldsymbol{\mathsf L}}\cdot\partial_{\bf X}p
\label{FP-eq-J}
\ee
expressed in terms of the matrix~(\ref{L}) of linear response coefficients,
and the rotational diffusion operator
\be
\hat{L}_{\rm r} p= \frac{D_{\rm r}}{\sin\theta} \Big\{ \partial_\theta\left[ \sin\theta \left( \partial_\theta p + \beta\mu B \sin\theta \, p\right)\right] + \frac{1}{\sin\theta}\, \partial_\phi^2 p\Big\} ,
\ee
where $D_{\rm r}\equiv k_{\rm B}T/\gamma_{\rm r}$ is the rotational diffusion coefficient.
We notice that $\hat{L}_{\rm r}{\cal P}_{\rm r}=0$ for the equilibrium rotational distribution
\be \label{rot-eq-distrib}
{\cal P}_{\rm r}({\bf u}) =  \frac{1}{Z_{\rm r}} \, \exp(-\beta U)
=  \frac{1}{Z_{\rm r}} \, \exp(\beta\mu B \, \cos\theta)
\ee
for the potential energy $U=-\mu {\bf B}\cdot{\bf u}$ of the magnetic moment in the external magnetic field ${\bf B}=(0,0,B)$ oriented in the $z$-direction.\cite{GM84}

If rotation is faster than reaction, we may suppose that
\be
p({\bf r},n,{\bf u};t) \simeq {\cal P}({\bf r},n;t) \, {\cal P}_{\rm r}({\bf u})
\ee
with the probability density
\be
{\cal P}({\bf r},n;t) = \int d\cos\theta \, d\phi \, p({\bf r},n,{\bf u};t)
\ee
governed by
\bea
\partial_t{\cal P} &=& - \left(\chi \, W_{\rm rxn} \, \langle{\bf u}\rangle + \beta D\, {\bf F}_{\rm ext}\right)\cdot\pmb{\nabla}{\cal P} \nonumber\\
&& - \left( W_{\rm rxn} + \beta\chi D_{\rm rxn} \langle{\bf u}\rangle\cdot{\bf F}_{\rm ext}\right) \partial_n{\cal P} \nonumber\\
&& + D\nabla^2 {\cal P} + 2  \chi \, D_{\rm rxn} \, \langle{\bf u}\rangle\cdot\pmb{\nabla}\partial_n{\cal P} + D_{\rm rxn}\partial_n^2{\cal P}  , \nonumber\\ &&
\eea
where $\langle\cdot\rangle$ denotes a statistical average with respect to the equilibrium canonical distribution~(\ref{rot-eq-distrib}) for the orientation.

If reaction is faster than rotation, the probability density
\be
{\mathscr P}({\bf r},{\bf u};t) = \int dn \, p({\bf r},n,{\bf u};t)
\ee
is ruled by
\be
\partial_t{\mathscr P} +\left(\chi \, W_{\rm rxn} \, {\bf u} + \beta D\, {\bf F}_{\rm ext}\right)\cdot\pmb{\nabla}{\mathscr P}= D\, \nabla^2{\mathscr P}  + \hat{L}_{\rm r} {\mathscr P} .
\ee
In the absence of external magnetic field ${\bf B}=0$, there is no preferential orientation so that the rotational motion remains diffusive and controlled by the rotational diffusion time
\be
{\mathscr T}_{\rm r} = \int_0^\infty dt \, \langle {\bf u}(0)\cdot{\bf u}(t)\rangle_{\rm eq} = \frac{1}{2 D_{\rm r}} .
\ee
In this case, we thus recover the known result~\cite{K13} that the effective diffusion coefficient is given by
$D_{\rm eff} = D + V_{\rm d}^2/(6 D_{\rm r})$. The larger the magnitude of the diffusiophoretic velocity $V_{\rm d}$, the more enhanced the diffusive random walk of the Janus particle.

A mechanochemical fluctuation theorem,\cite{GK17} which is a consequence of microreversibility,  was derived for the nonequilibrium process ruled by the Fokker-Planck equation~(\ref{FP-eq}) with~(\ref{FP-eq-J}).

\subsection{Mechanochemical coupling and efficiencies}

\begin{figure}[h]
\centerline{\scalebox{0.45}{\includegraphics{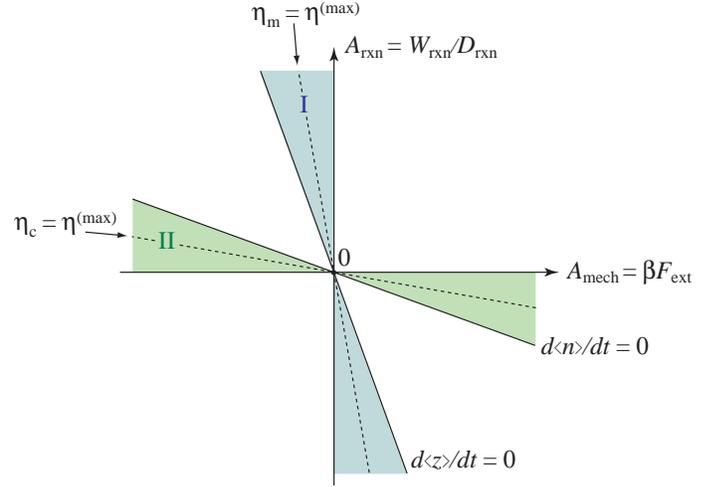}}}
\caption{Schematic representation of the different regimes of the active particle in the plane of the mechanical and chemical affinities.}
\label{fig}
\end{figure}

In order to investigate the implications of the previous results, we suppose that the external force and the magnetic field are oriented in the $z$-direction, so that ${\bf F}_{\rm ext}=(0,0,F)$ and ${\bf B}=(0,0,B)$.  Therefore, the particle is oriented on average in the same direction: $\langle u_z\rangle={\rm coth}(\beta\mu B)-1/(\beta\mu B)$.  In this case, the averages of Eqs.~(\ref{eq-r-fin}) and~(\ref{eq-n-fin}) give
\bea
\frac{d\langle z\rangle}{dt} &=& \chi  W_{\rm rxn}  \langle u_z\rangle + \beta D F \, , \label{eq-z-av}\\
\frac{d\langle n\rangle}{dt} &=& W_{\rm rxn} + \beta\chi D_{\rm rxn} \langle u_z\rangle F \, , \label{eq-n-av}
\eea
$d\langle x\rangle/dt=0$, and $d\langle y\rangle/dt=0$.  Depending on the values of the mechanical and chemical affinities the mean velocity $\langle\dot z\rangle$ and rate $\langle\dot n\rangle$ can take positive, vanishing, or negative values.  Equation~(\ref{eq-z-av}) shows that the mean velocity vanishes at the stall force $F_{\rm stall}=-F_{\rm d}\langle u_z\rangle$, which is proportional to the diffusiophoretic force $F_{\rm d}=\gamma\chi W_{\rm rxn}$. Also, the mean reaction rate is equal to zero at the force $F_0=-W_{\rm rxn}/(\beta\chi D_{\rm rxn}\langle u_z\rangle)$ according to Eq.~(\ref{eq-n-av}).  These two conditions are depicted in Fig.~\ref{fig} that shows the plane of the mechanical and chemical affinities.  For positive values of the chemical affinity $A_{\rm rxn}$, the propulsion driven by the reaction exerts a mechanical work if the force is in the range $F_{\rm stall} <F<0$, corresponding to the domain I in Fig.~\ref{fig}.  If the force is sufficiently opposed to propulsion to satisfy $F<F_0$, the mean reaction rate~(\ref{eq-n-av}) can become negative, meaning that the reaction is reversed and product is synthesized from reactants, instead of being consumed.

As for molecular motors,\cite{JAP97} the efficiency of the mechanical power of the motor can be characterized by
\be
\eta_{\rm m}\equiv-\frac{A_{\rm mech}\langle\dot z\rangle}{A_{\rm rxn}\langle \dot n\rangle} ,
\ee
and the efficiency of the reverse process of synthesis by $\eta_{\rm c}\equiv 1/\eta_{\rm m}$.
Since the thermodynamic entropy production rate of the coupled processes
\be
\frac{1}{k_{\rm B}}\frac{d_{\rm i}S}{dt} =A_{\rm mech}\langle{\dot z}\rangle + A_{\rm rxn}\langle\dot n\rangle \geq 0
\label{entr-prod}
\ee
is non-negative according to the second law of thermodynamics, the mechanical and chemical efficiencies are bounded by $0\leq\eta_{\rm m}\leq 1$ and $0\leq\eta_{\rm c}\leq 1$ in their respective domains of application.  For a given chemical affinity, the maximal value of the mechanical efficiency is given by
\be
\eta^{\rm (max)}=\frac{1-\sqrt{1-q^2}}{1+\sqrt{1-q^2}}=\frac{q^2}{4}+\frac{q^4}{8}+O(q^6) \, ,
\ee
where $q\equiv\chi\langle u_z\rangle\sqrt{D_{\rm rxn}/D}$
 satisfies the condition $q^2\leq 1$.  Accordingly, the efficiency of self-propulsion increases with the diffusiophoretic coupling $\vert\chi\vert$.  A similar expression holds for the chemical efficiency.\cite{GK17}  The locations where the efficiencies reach their maximal values are depicted as dashed lines in Fig.~\ref{fig}.  These results show the analogy between self-diffusiophoretic active particles and molecular motors.

\section{Conclusion and perspectives}
\label{conclusion}

Equations of motion describing the stochastic dynamics and reaction of an active Janus particle self-propelled by diffusiophoresis were derived in this paper, starting from the fluctuating chemohydrodynamics of the solution surrounding the particle and the boundary conditions for the fluid velocity and concentration fields at the interface with the particle. Utilizing Green-function methods and generalizations of the F\'axen theorem,\cite{MB74,BM74,ABM75,BAM77,LK79} the frequency-dependent force, torque, and reaction rate of the particle were deduced from the boundary conditions and the fluctuating chemohydrodynamic equations.  In this way, the contributions to the force and torque from friction by the fluid viscosity, diffusiophoresis by the concentration gradients self-generated by the reaction, and the possible presence of an external force and an external torque were determined. From these general equations, coupled overdamped Langevin equations were deduced for the translation, rotation, and reaction of the Janus particle.  The stochastic equation for reaction includes a contribution due to mechanochemical coupling that is required in order to satisfy the Onsager-Casimir reciprocal relations to be consistent with microreversibility.  This contribution is essential to establish the previously derived mechanochemical fluctuation theorem.\cite{GK17}  The Fokker-Planck equation associated with the coupled stochastic equations was analyzed in several limiting situations.  Moreover, the implications of the mechanochemical coupling were studied, in particular, for the efficiencies of energy transduction.

The results obtained in this paper provide fresh perspectives on the understanding of self-propulsion mechanisms of active particles. The calculations can be extended easily to treat other self-phoretic mechanisms including electrophoresis and thermophoresis.  For thermophoresis, the coupling of boundary conditions describing thermal slip are known~\cite{W67,BAM76} and the methods we have developed can be applied {\it mutatis mutandis} to this other phoretic mechanism.  For electrophoresis, the description should include the electric field generated by the electric charges of the particle and the electrolytic solution.

One can also consider reactions that are more complex than the simple ${\rm A}\rightleftharpoons{\rm B}$ reaction that was used to illustrate our results.  Indeed, the experimental systems that have been studied typically involve nonlinear reaction networks with possible adsorbates at the surface of the active particle.  Accordingly, the surface densities of these adsorbates should be included in the description. The motion of active particles with nonspherical shapes can also be studied.  For such active particles, nontrivial coupling between translational and rotational motions are expected in Eqs.~(\ref{Langevin-eq-transl}) and~(\ref{Langevin-eq-rot}).

Furthermore, one can envisage situations where the medium surrounding the active particle is a rarefied or dilute gas, instead of a liquid solution.  Such systems may be described by the Boltzmann equation, instead of the Navier-Stokes and diffusion equations.  In rarefied gases, the mean free path is larger than the particle radius so that the reactants are essentially in free flight before colliding and reacting at the particle surface.  In this case, the transport can be supposed to be ballistic in the gas, as for surface reactions studied in ultra-high vacuum conditions.  In dilute gases, the mean free path is smaller than the particle radius so that a local equilibrium will establish itself on the mean free path length scale and intercollisional time scale, whereupon transport is no longer ballistic.  The effects of long-time tails on diffusiophoresis could be investigated in such situations.\cite{LR13}

%%%%%%%%%%%%%%%%%%%%%%%%%%%%%%%%%%%%%%%%%%%%%%%%%
\section*{Acknowledgments}

The Authors thank Patrick Grosfils and Mu-Jie Huang for fruitful discussions. Financial support from the International Solvay Institutes for Physics and Chemistry, the Universit\'e libre de Bruxelles (ULB), the Fonds de la Recherche Scientifique~-~FNRS under the Grant PDR~T.0094.16 for the project ``SYMSTATPHYS", and the Natural Sciences and Engineering Research Council of Canada is acknowledged.

%%%%%%%%%%%%%%%%%%%%%%%%%%%%%%%%%%%%%%%%%%%%%%%%%

\appendix
\section{Solution of reaction-diffusion equation}
\label{AppA}

Although a set of linear equations must be solved numerically to obtain the concentration fields, insight into the nature of these fields can be obtained by considering the reaction-limited and diffusion-limited regimes.

Since the matrix ${\boldsymbol{\mathsf M}}$ in Sec.~\ref{sph-J-part} is the sum of a diagonal matrix and a square matrix multiplied by ${\rm Da}$, in the reaction-limited regime it may be approximated by an expansion in powers of ${\rm Da}$. The coefficient $a_1$ takes the value $a_1= 3/8=0.375$ for ${\rm Da}=0$. In this limit we also have $\gamma_J=0.708115$.

\begin{figure}[h]
\centerline{\scalebox{0.5}{\includegraphics{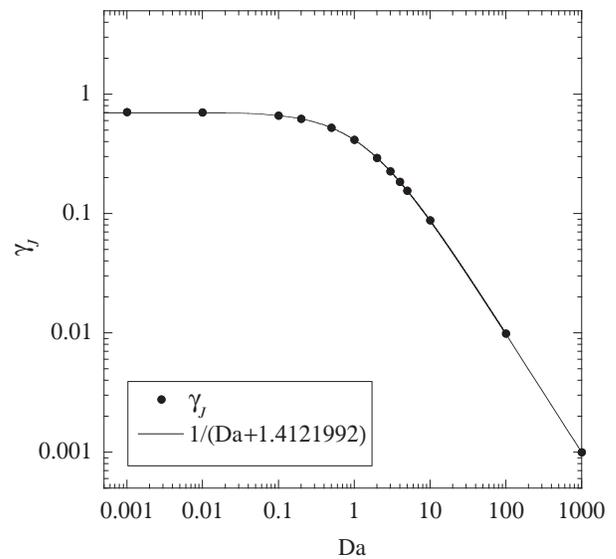}}}
\caption{The coefficient $\gamma_J$ versus the parameter ${\rm Da}$ computed with the expansion~(\ref{f}) truncated at $l\leq 30$ (dots) and compared with the fit (line) given by Eq.~(\ref{gamma_J_fit}).}
\label{figA}
\end{figure}

In the diffusion-limited regime, there is a thin and strong depletion zone close to the catalytic surface.  Accordingly, the problem can be approximately solved by considering a nearly flat catalytic surface.  Close to the catalytic surface, the solution is given by
\be
f(r,\theta) \simeq \frac{1}{{\rm Da}+1} \, \frac{R}{r} \, ,
\ee
while $f\simeq 0$ close to the noncatalytic surface. The solution switches from one form to the other at the contact line between the catalytic and noncatalytic hemispheres, but the contribution of this line is negligible because ${\rm Da}\gg 1$.  Thus, we have
\be
f(R,\theta)=\left\{\begin{array}{lr}
({\rm Da}+1)^{-1}, &\text{on catalytic surface,} \\
\, 0\, , &\text{on noncatalytic surface.}
\end{array}\right.
\ee
Over the same range of ${\rm Da}$ values, we have that $a_1\simeq 0.75\, {\rm Da}^{-1}$ and $\gamma_J\simeq {\rm Da}^{-1}$ for ${\rm Da}\to \infty$. Combining the behavior in both regimes, the coefficient $a_1$ can be fit by $a_1 \simeq 0.75({\rm Da}+2)^{-1}$. Similarly, we find that $\gamma_J$ can be approximated by
\be
\gamma_J \simeq ({\rm Da}+1.4121992)^{-1} , \label{gamma_J_fit}
\ee
at the crossover between the reaction- and diffusion-limited regimes, as shown in Fig.~\ref{figA}.

%%%%%%%%%%%%%%%%%%%%%%%%%%%%%%%%%%%%%%%%%%%%%%%%

%

\end{document}